%% file: main.tex
\documentclass[submission]{eptcs}
\usepackage{prelude}
\input{macros}
\title{Solving the TTC 2011 Compiler Optimization Case with GROOVE}%
\author{Arend Rensink%
\qquad\qquad%
Eduardo Zambon%
\institute{%
Department of Computer Science\\%
University of Twente, The Netherlands%
}%
\email{\{rensink, zambon\}@cs.utwente.nl}%
}%
\def\titlerunning{Solving the TTC 2011 Compiler Optimization Case with GROOVE}%
\def\authorrunning{A. Rensink \& E. Zambon}%
\begin{document}
\maketitle
\begin{abstract}
This report presents a partial solution to the Compiler Optimization case study
\cite{compileroptimizationcase}
using \GROOVE. We explain how the input graphs provided with the case study were
adapted into a \GROOVE representation and we describe an initial solution for
Task 1. This solution allows us to automatically reproduce the steps of the
constant folding example given in the case description. We did not solve Task 2.
\end{abstract}
%
%
\section{GROOVE}
\seclabel{groove}
\GROOVE\footnote{Available at \url{http://groove.cs.utwente.nl}} \cite{sttt}
is a general purpose graph transformation tool set that uses simple labelled
graphs. The core functionality of \GROOVE is to recursively apply all rules from
a predefined set (the graph production system -- GPS) to a given start graph,
and to all graphs generated by such applications. This results in a \emph{state
space} consisting of the generated graphs.
The main component of the \GROOVE tool set is the Simulator, a graphical tool
that integrates (among others) the functionalities of rule and host graph
editing, and of interactive or automatic state space exploration.

\subsection{Host Graphs}

In \GROOVE, the host graphs, i.e., the graphs to be transformed, are simple
graphs with nodes and directed labelled edges. In simple graphs, edges do not
have an identity, and therefore parallel edges (i.e., edges with same label, and
source and target nodes) are not allowed. Also, for the same reason, edges may
not have attributes.
In the graphical representation, nodes are depicted as rectangles and edges as
binary arrows between two nodes. Node labels can be either node types or flags.
Node types [resp. flags] are displayed in {\bf bold} [resp. {\it italic}] inside
a node rectangle.

\subsection{Rules}
The transformation rules in \GROOVE are represented by a single graph and
colours and shapes are used to distinguish different elements.
\figref{example-rule} shows a small example rule.
\begin{itemize}
\item {\bf Readers.} The black (continuous thin) nodes and edges must be present
in the host graph for the rule to be applicable and are preserved by the rule
application;
\item {\bf Embargoes.} The red (dashed fat) nodes and edges must be absent in
the host graph for the rule to be applicable;
\item {\bf Erasers.} The blue (dashed thin) nodes and edges must be present in
the host graph for the rule to be applicable and are deleted by the rule
application;
\item {\bf Creators.} The green (continuous fat) nodes and edges are created by
the rule application.
\end{itemize}
\begin{figure}
\centering
\scalebox{\mytikzscale}{\begin{tabular}[t]{@{}l@{}}\input{figs/example-rule.tikz}\end{tabular}}%
 \qquad
\begin{tabular}[b]{ll}
\multicolumn{2}{l}{\it Legend:} \\
\scalebox{\mytikzscale}{\begin{tikzpicture}[scale=\tikzscale]
  \node[node] (n0) {\ml{\textit{A}}};
  \node[node] (n1) [right=of n0] {\ml{\textit{A}}};
  \path (n0) edge[edge] node[lab] {b} (n1);
\end{tikzpicture}} & Matched and preserved \\
\scalebox{\mytikzscale}{\begin{tikzpicture}[scale=\tikzscale]
  \node[node] (n0) {\ml{\textit{A}}};
  \node[nacnode] (n1) [right=of n0] {\ml{\textit{A}}};
  \path (n0) edge[nacedge] node[naclab] {b} (n1);
\end{tikzpicture}} & Forbidden \\
\scalebox{\mytikzscale}{\begin{tikzpicture}[scale=\tikzscale]
  \node[node] (n0) {\ml{\textit{A}}};
  \node[delnode] (n1) [right=of n0] {\ml{\textit{A}}};
  \path (n0) edge[deledge] node[dellab] {b} (n1);
\end{tikzpicture}} & Matched and deleted \\
\scalebox{\mytikzscale}{\begin{tikzpicture}[scale=\tikzscale]
  \node[node] (n0) {\ml{\textit{A}}};
  \node[newnode] (n1) [right=of n0] {\ml{\textit{A}}};
  \path (n0) edge[newedge] node[newlab] {b} (n1);
\end{tikzpicture}} & Created \\
\end{tabular}
\caption{Example \GROOVE rule and legend}
\figlabel{example-rule}
\end{figure}
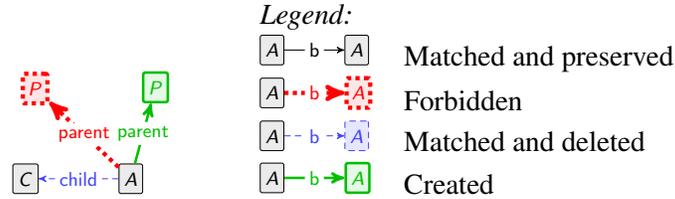
Embargo elements are usually called Negative Application Conditions (NACs).
When a node type or flag is used in a non-reader element but the node itself is
not modified, the node type or flag is prefixed with a character to indicate its
role. The characters used are $+$, $-$, and $!$, for creator, eraser, and
embargo elements, respectively.
%
%
%
\section{Solution}
\seclabel{solution}
%

\subsection{Input from the \FIRM representation}

The input graphs provided with the case study are stored in GXL format and
conform to the \FIRM representation. \GROOVE also uses GXL to store graphs but
it was not possible to immediately load the given files because the input graphs
have certain properties that are not compatible with \GROOVE (e.g., edges with
attributes) and therefore require some adaptation.

The case description lists all node and edge types that may occur in the program
graphs. These types are also included in the GXL files given on the form of a
type graph. Based on these two sources of information we constructed our own
type graph\footnote{\GROOVE enforces static typing, so there is no overhead for
type checking while performing a transformation. Using a type graph is a very
convenient way to avoid simple mistakes (e.g., typos) while creating a grammar.}
in \GROOVE, shown in \figref{type}.

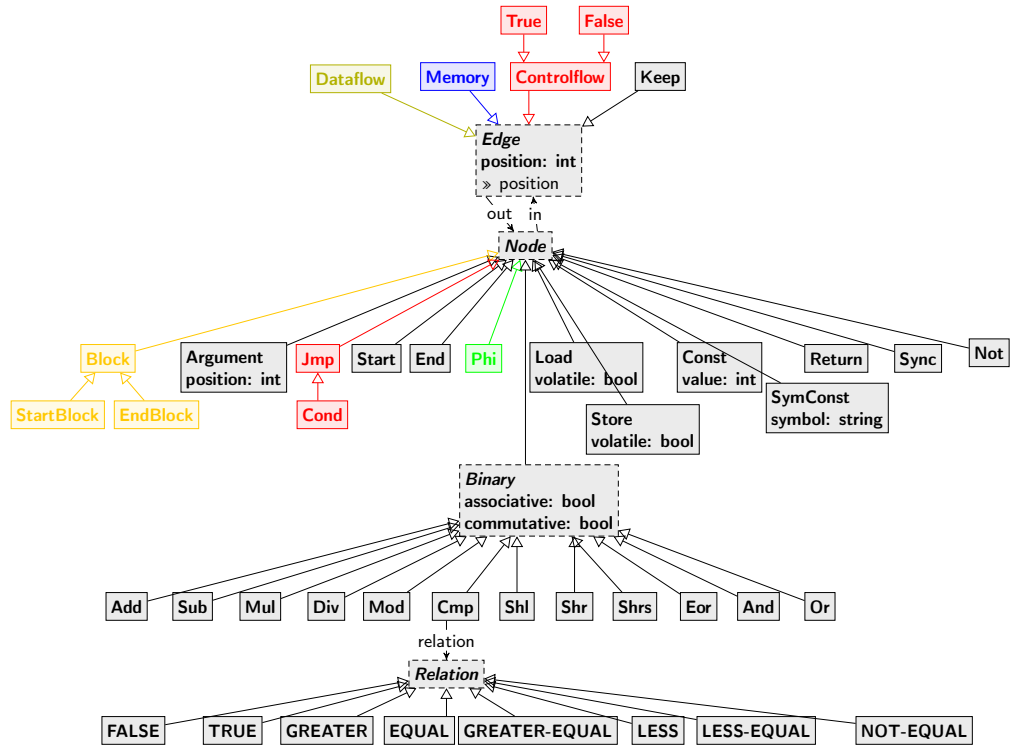
\begin{figure}
\centering
\scalebox{0.75}{\begin{tabular}[t]{@{}l@{}}\input{figs/type.tikz}\end{tabular}}%

\caption{Type graph for the adapted input graphs.}
\figlabel{type}
\end{figure}

Each node in the figure corresponds to a node type; some have associated
attributes. Types shown in \textbf{\textit{bold italic}} inside dashed nodes are
abstract. Edges with open triangular arrows indicate type inheritance. A key
point in the type graph shown in \figref{type} is the following. In \GROOVE
edges do not have types or attributes while in \FIRM the edges do. To encode
these extra properties in \GROOVE, edges have to be \emph{nodified}, i.e., each
edge of the \FIRM graph is transformed to a node in the \GROOVE graph with a
proper sub-type of \Edge and associated \position attribute. Nodes representing
operations, i.e., sub-types of \Node, are connected via \Edge nodes and
associated edges labelled {\sf in} and {\sf out}. The remaining elements of the
type graph of \figref{type} correspond directly to the ones described in the
case study. 

After creating our type graph, the program graph used in the constant folding
example was manually created by inspecting the given GXL file and the
corresponding figure in the case study. Our start graph in a plain
representation, i.e., without block containment visualisation, is shown in
\figref{start}.

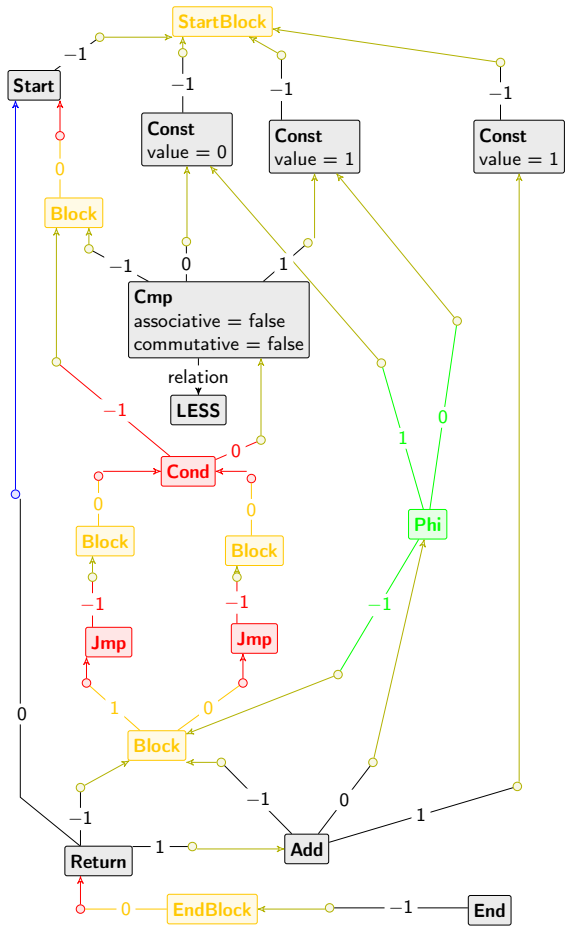
\begin{figure}
\centering
\scalebox{\mytikzscale}{\begin{tabular}[t]{@{}l@{}}\input{figs/start.tikz}\end{tabular}}%

\caption{Program graph of minimum plus one function with constants.}
\figlabel{start}
\end{figure}

We would like to point that, despite the current manual adaptation of the input
graphs, there are no technical limitations that prevent the automatic loading
of \FIRM graphs using the conversion described above. Automatic loading was not
done due to time limitations only.

\subsection{Verifier}

We implemented the sanity checks described in the case study in negated form
such that if an invalid configuration is produced, then a checking rule matches.
\figref{consts} shows rule {\sf consts}, that is triggered if there is a
constant located in a \Block that is not the \StartBlock. The other checks given
in the case study were implemented with the rules named {\sf single-start}, {\sf
single-end}, {\sf containment}, {\sf phi-check}, and {\sf pos-check}. (See the
grammar for the solution in the SHARE image \cite{share}.)

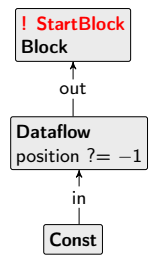
\begin{figure}
\centering
\scalebox{\mytikzscale}{\begin{tabular}[t]{@{}l@{}}\input{figs/consts.tikz}\end{tabular}}%

\caption{Sanity check rule {\sf consts}.}
\figlabel{consts}
\end{figure}

\subsection{Constant Folding}

To solve the constant folding example given in the case description we created
seven rules to perform the folding of operations and another three cleanup
rules to handle dangling edges and constants without references.

\figref{add-fold} shows rule {\sf add-fold-int}, that performs the last step of
the transformation: folding of an \Add operation with two constant operands. The
\Add node and the two \Dataflow edges associated with the operands are deleted
by the rule. The constants used in the addition are not removed because they
may be referenced by other operations. A new constant is created with the
result of the addition and all \Dataflow edges incoming into \Add are re-routed
to the newly created constant. We use a special quantifier node (labelled with
$\forall^{>0}$) to redirect an arbitrary number of \Dataflow edges.

\begin{figure}
\centering
\scalebox{\mytikzscale}{\begin{tabular}[t]{@{}l@{}}\input{figs/add-fold.tikz}\end{tabular}}%

\caption{Rule {\sf add-fold-int}, for folding the addition of two integer
constants.}
\figlabel{add-fold}
\end{figure}
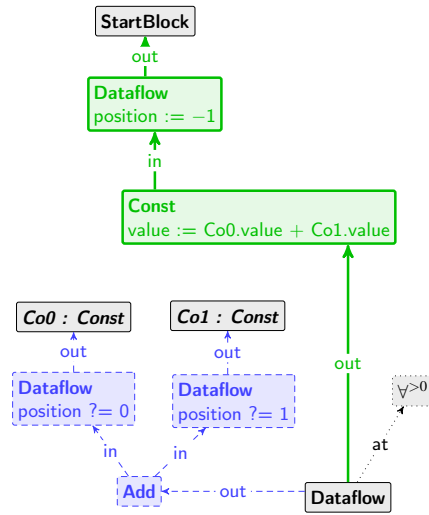

The remaining rules are similar. We created one rule for each operation
folding, except for the handling of unreachable blocks, which uses two rules,
one for removing blocks and another for adjusting the edges of {\sf Phi} nodes.
(Again, for the complete solution, we refer the reader to the grammar that is
available in the SHARE image \cite{share}.)

We did not create folding rules for the operations that do not occur in the
given example. Still, once more, we do not foresee any technical difficulties
to do so. The remaining operations were not handled only due to limited time
availability.

%
%
\section{Conclusion}
\seclabel{conclusion}
In this report we presented the key points of our solution for Task 1 of the
case study. Task 2 was not addressed. We conclude with an overview according
to the criteria listed in the case study.

\begin{itemize}
\item {\bf Completeness.} Since we did not cover all operations, it is expected
that no program graph other than the example discussed can be handled. Absence
of automatic loading of graphs is another limitation that prevents the use of
the test suite.
\item {\bf Performance.} N/A. See reasons in the item above.
\item {\bf Conciseness.} As a rule of thumb, we have one rule for each
operation.
\item {\bf Purity.} The solution is entirely made of graph transformations, no
glue code is necessary.
\end{itemize}

%
%
\bibliographystyle{eptcs}
\bibliography{main}
\end{document}

%% file: macros.tex
\newcommand{\GROOVE} {{\sc groove}\xspace}
\newcommand{\FIRM} {{\sc Firm}\xspace}

\newcommand{\Node} {{\sf\textbf{\textit{Node}}}\xspace}
\newcommand{\Edge} {{\sf\textbf{\textit{Edge}}}\xspace}
\newcommand{\position} {{\sf position}\xspace}
\newcommand{\Block} {{\sf Block}\xspace}
\newcommand{\StartBlock} {{\sf StartBlock}\xspace}
\newcommand{\Add} {{\sf Add}\xspace}
\newcommand{\Dataflow} {{\sf Dataflow}\xspace}

%% file: figs/example-rule.tikz
\begin{tikzpicture}[scale=\tikzscale]
\node[node] (n3)  at (1.400, -1.265) {\ml{\textit{C}}};
\node[nacnode] (n2)  at (1.480, -0.515) {\ml{\textit{P}}};
\node[newnode] (n1)  at (2.480, -0.505) {\ml{\textit{P}}};
\node[node] (n0)  at (2.275, -1.265) {\ml{\textit{A}}};
\path[newedge] (n0)  -- node[newlab]{parent} (n1) ;
\path[nacedge] (n0)  -- node[naclab]{parent} (n2) ;
\path[deledge](n0.west |- 1.400, -1.265) -- node[dellab]{child} (n3) ;
\userdefinedmacro
\end{tikzpicture}
\renewcommand{\userdefinedmacro}{\relax}

%% file: figs/type.tikz
%
\definecolor{n8c}{RGB}{255,0,0}
\definecolor{n43c}{RGB}{0,0,255}
\definecolor{n0c}{RGB}{255,200,0}
\definecolor{n7c}{RGB}{0,255,0}
\definecolor{n1c}{RGB}{255,200,0}
\definecolor{n9c}{RGB}{255,0,0}
\definecolor{n2c}{RGB}{255,200,0}
\definecolor{n44c}{RGB}{255,0,0}
\definecolor{n46c}{RGB}{255,0,0}
\definecolor{n45c}{RGB}{255,0,0}
\definecolor{n42c}{RGB}{178,178,0}
\begin{tikzpicture}[
n8cs/.style={draw=n8c,text=n8c,fill=n8c!10},
n43cs/.style={draw=n43c,text=n43c,fill=n43c!10},
n0cs/.style={draw=n0c,text=n0c,fill=n0c!10},
n7cs/.style={draw=n7c,text=n7c,fill=n7c!10},
n1cs/.style={draw=n1c,text=n1c,fill=n1c!10},
n9cs/.style={draw=n9c,text=n9c,fill=n9c!10},
n2cs/.style={draw=n2c,text=n2c,fill=n2c!10},
n44cs/.style={draw=n44c,text=n44c,fill=n44c!10},
n46cs/.style={draw=n46c,text=n46c,fill=n46c!10},
n45cs/.style={draw=n45c,text=n45c,fill=n45c!10},
n42cs/.style={draw=n42c,text=n42c,fill=n42c!10},
scale=\tikzscale]
\node[node, type] (n27)  at (5.845, -5.460) {\ml{\textbf{Shrs}}};
\node[node, type] (n10)  at (7.630, -3.260) {\ml{\textbf{Return}}};
\node[node, type, n44cs] (n44)  at (5.185, -0.760) {\ml{\textbf{Controlflow}}};
\node[node, type] (n4)  at (4.030, -3.260) {\ml{\textbf{End}}};
\node[node, type, n1cs] (n1)  at (0.730, -3.760) {\ml{\textbf{StartBlock}}};
\node[node, type] (n12)  at (8.355, -3.260) {\ml{\textbf{Sync}}};
\node[node, type] (n38)  at (4.985, -6.560) {\ml{\textbf{GREATER-EQUAL}}};
\node[node, type, n42cs] (n42)  at (3.325, -0.780) {\ml{\textbf{Dataflow}}};
\node[node, type] (n31)  at (4.255, -5.460) {\ml{\textbf{Cmp}}};
\node[node, type] (n37)  at (3.935, -6.560) {\ml{\textbf{EQUAL}}};
\node[node, type] (n30)  at (6.410, -5.460) {\ml{\textbf{Eor}}};
\node[node, type, n45cs] (n45)  at (4.860, -0.260) {\ml{\textbf{True}}};
\node[node, type] (n22)  at (2.525, -5.460) {\ml{\textbf{Mul}}};
\node[node, type] (n34)  at (1.400, -6.560) {\ml{\textbf{FALSE}}};
\node[node, type] (n5)  at (2.290, -3.340) {\ml{\textbf{Argument}\\\textbf{position: int}}};
\node[node, type] (n18)  at (8.985, -3.210) {\ml{\textbf{Not}}};
\node[node, type] (n13)  at (7.550, -3.700) {\ml{\textbf{SymConst}\\\textbf{symbol: string}}};
\node[node, type] (n39)  at (6.050, -6.560) {\ml{\textbf{LESS}}};
\node[node, type, n2cs] (n2)  at (1.605, -3.760) {\ml{\textbf{EndBlock}}};
\node[node, type, n43cs] (n43)  at (4.275, -0.760) {\ml{\textbf{Memory}}};
\node[node, type] (n20)  at (1.340, -5.460) {\ml{\textbf{Add}}};
\node[node, type] (n35)  at (2.280, -6.560) {\ml{\textbf{TRUE}}};
\node[absnode, type] (n19)  at (4.995, -4.520) {\ml{\textbf{\textit{Binary}}\\\textbf{associative: bool}\\\textbf{commutative: bool}}};
\node[node, type, n7cs] (n7)  at (4.505, -3.260) {\ml{\textbf{Phi}}};
\node[node, type, n0cs] (n0)  at (1.175, -3.260) {\ml{\textbf{Block}}};
\node[node, type] (n40)  at (6.920, -6.560) {\ml{\textbf{LESS-EQUAL}}};
\node[node, type] (n17)  at (5.915, -3.890) {\ml{\textbf{Store}\\\textbf{volatile: bool}}};
\node[node, type] (n41)  at (8.320, -6.560) {\ml{\textbf{NOT-EQUAL}}};
\node[absnode, type] (n33)  at (4.175, -6.060) {\ml{\textbf{\textit{Relation}}}};
\node[node, type, n8cs] (n8)  at (3.035, -3.260) {\ml{\textbf{Jmp}}};
\node[absnode, type] (n32)  at (4.905, -1.505) {\ml{\textbf{\textit{Edge}}\\\textbf{position: int}\\>> position}};
\node[absnode, type] (n49)  at (4.875, -2.260) {\ml{\textbf{\textit{Node}}}};
\node[node, type] (n47)  at (6.070, -0.760) {\ml{\textbf{Keep}}};
\node[node, type] (n15)  at (5.415, -3.340) {\ml{\textbf{Load}\\\textbf{volatile: bool}}};
\node[node, type] (n26)  at (5.310, -5.460) {\ml{\textbf{Shr}}};
\node[node, type] (n28)  at (6.940, -5.460) {\ml{\textbf{And}}};
\node[node, type] (n23)  at (3.110, -5.460) {\ml{\textbf{Div}}};
\node[node, type] (n24)  at (3.645, -5.460) {\ml{\textbf{Mod}}};
\node[node, type, n46cs] (n46)  at (5.570, -0.260) {\ml{\textbf{False}}};
\node[node, type] (n25)  at (4.800, -5.460) {\ml{\textbf{Shl}}};
\node[node, type] (n21)  at (1.925, -5.460) {\ml{\textbf{Sub}}};
\node[node, type, n9cs] (n9)  at (3.075, -3.760) {\ml{\textbf{Cond}}};
\node[node, type] (n36)  at (3.115, -6.560) {\ml{\textbf{GREATER}}};
\node[node, type] (n3)  at (3.555, -3.260) {\ml{\textbf{Start}}};
\node[node, type] (n11)  at (6.595, -3.340) {\ml{\textbf{Const}\\\textbf{value: int}}};
\node[node, type] (n29)  at (7.485, -5.460) {\ml{\textbf{Or}}};
\path[subedge] (n18)  --  (n49) ;
\path[subedge] (n24)  --  (n19) ;
\path[subedge, n1cs] (n1)  --  (n0) ;
\path[subedge, n7cs] (n7)  --  (n49) ;
\path[subedge] (n3)  --  (n49) ;
\path[subedge](n37.north -| 4.175, -6.060) --  (n33) ;
\path[subedge] (n47)  --  (n32) ;
\path[subedge](n19.north -| 4.875, -2.260) --  (n49) ;
\path[edge] (n49.50)  -- node[lab]{in} (n32);
\path[subedge] (n5)  --  (n49) ;
\path[subedge] (n31)  --  (n19) ;
\path[subedge] (n39)  --  (n33) ;
\path[subedge] (n34)  --  (n33) ;
\path[subedge] (n26)  -- (n19.south -| 5.310, -5.460);
\path[subedge, n0cs] (n0)  --  (n49) ;
\path[subedge] (n29)  --  (n19) ;
\path[subedge, n2cs] (n2)  --  (n0) ;
\path[subedge] (n21)  --  (n19) ;
\path[subedge, n8cs] (n8)  --  (n49) ;
\path[edge](n31.south -| 4.175, -6.060) -- node[lab]{relation} (n33) ;
\path[subedge] (n40)  --  (n33) ;
\path[subedge, n44cs](n44.south -| 4.905, -1.505) --  (n32) ;
\path[subedge] (n15)  --  (n49) ;
\path[subedge, n9cs](n9.north -| 3.035, -3.260) --  (n8) ;
\path[subedge, n42cs] (n42)  --  (n32) ;
\path[subedge] (n4)  --  (n49) ;
\path[subedge, n45cs] (n45)  -- (n44.north -| 4.860, -0.260);
\path[subedge] (n17)  --  (n49) ;
\path[subedge] (n25)  -- (n19.south -| 4.800, -5.460);
\path[subedge] (n36)  --  (n33) ;
\path[subedge] (n20)  --  (n19) ;
\path[subedge, n46cs] (n46)  -- (n44.north -| 5.570, -0.260);
\path[subedge] (n30)  --  (n19) ;
\path[subedge, n43cs] (n43)  --  (n32) ;
\path[subedge] (n41)  --  (n33) ;
\path[subedge] (n23)  --  (n19) ;
\path[subedge] (n12)  --  (n49) ;
\path[subedge] (n38)  --  (n33) ;
\path[subedge] (n22)  --  (n19) ;
\path[subedge] (n27)  --  (n19) ;
\path[subedge] (n13)  --  (n49) ;
\path[subedge] (n10)  --  (n49) ;
\path[subedge] (n35)  --  (n33) ;
\path[edge](n32.220) -- node[lab]{out} (n49);
\path[subedge] (n28)  --  (n19) ;
\path[subedge] (n11)  --  (n49) ;
\userdefinedmacro
\end{tikzpicture}
\renewcommand{\userdefinedmacro}{\relax}

%% file: figs/start.tikz
%
\definecolor{n23c}{RGB}{178,178,0}
\definecolor{n51c}{RGB}{255,0,0}
\definecolor{n12c}{RGB}{255,0,0}
\definecolor{n44c}{RGB}{255,0,0}
\definecolor{n31c}{RGB}{178,178,0}
\definecolor{n14c}{RGB}{255,200,0}
\definecolor{n49c}{RGB}{178,178,0}
\definecolor{n77c}{RGB}{178,178,0}
\definecolor{n65c}{RGB}{178,178,0}
\definecolor{n47c}{RGB}{178,178,0}
\definecolor{n59c}{RGB}{178,178,0}
\definecolor{n67c}{RGB}{178,178,0}
\definecolor{n42c}{RGB}{178,178,0}
\definecolor{n3c}{RGB}{255,200,0}
\definecolor{n46c}{RGB}{255,0,0}
\definecolor{n62c}{RGB}{178,178,0}
\definecolor{n15c}{RGB}{0,255,0}
\definecolor{n4c}{RGB}{255,200,0}
\definecolor{n0c}{RGB}{255,200,0}
\definecolor{n57c}{RGB}{178,178,0}
\definecolor{n11c}{RGB}{255,0,0}
\definecolor{n26c}{RGB}{255,0,0}
\definecolor{n33c}{RGB}{178,178,0}
\definecolor{n36c}{RGB}{178,178,0}
\definecolor{n28c}{RGB}{178,178,0}
\definecolor{n37c}{RGB}{0,0,255}
\definecolor{n2c}{RGB}{255,200,0}
\definecolor{n75c}{RGB}{178,178,0}
\definecolor{n21c}{RGB}{255,200,0}
\definecolor{n53c}{RGB}{255,0,0}
\definecolor{n30c}{RGB}{178,178,0}
\definecolor{n55c}{RGB}{178,178,0}
\definecolor{n40c}{RGB}{178,178,0}
\definecolor{n79c}{RGB}{178,178,0}
\definecolor{n64c}{RGB}{178,178,0}
\definecolor{n70c}{RGB}{255,0,0}
\definecolor{n13c}{RGB}{255,0,0}
\begin{tikzpicture}[
n23cs/.style={draw=n23c,text=n23c,fill=n23c!10},
n51cs/.style={draw=n51c,text=n51c,fill=n51c!10},
n12cs/.style={draw=n12c,text=n12c,fill=n12c!10},
n44cs/.style={draw=n44c,text=n44c,fill=n44c!10},
n31cs/.style={draw=n31c,text=n31c,fill=n31c!10},
n14cs/.style={draw=n14c,text=n14c,fill=n14c!10},
n49cs/.style={draw=n49c,text=n49c,fill=n49c!10},
n77cs/.style={draw=n77c,text=n77c,fill=n77c!10},
n65cs/.style={draw=n65c,text=n65c,fill=n65c!10},
n47cs/.style={draw=n47c,text=n47c,fill=n47c!10},
n59cs/.style={draw=n59c,text=n59c,fill=n59c!10},
n67cs/.style={draw=n67c,text=n67c,fill=n67c!10},
n42cs/.style={draw=n42c,text=n42c,fill=n42c!10},
n3cs/.style={draw=n3c,text=n3c,fill=n3c!10},
n46cs/.style={draw=n46c,text=n46c,fill=n46c!10},
n62cs/.style={draw=n62c,text=n62c,fill=n62c!10},
n15cs/.style={draw=n15c,text=n15c,fill=n15c!10},
n4cs/.style={draw=n4c,text=n4c,fill=n4c!10},
n0cs/.style={draw=n0c,text=n0c,fill=n0c!10},
n57cs/.style={draw=n57c,text=n57c,fill=n57c!10},
n11cs/.style={draw=n11c,text=n11c,fill=n11c!10},
n26cs/.style={draw=n26c,text=n26c,fill=n26c!10},
n33cs/.style={draw=n33c,text=n33c,fill=n33c!10},
n36cs/.style={draw=n36c,text=n36c,fill=n36c!10},
n28cs/.style={draw=n28c,text=n28c,fill=n28c!10},
n37cs/.style={draw=n37c,text=n37c,fill=n37c!10},
n2cs/.style={draw=n2c,text=n2c,fill=n2c!10},
n75cs/.style={draw=n75c,text=n75c,fill=n75c!10},
n21cs/.style={draw=n21c,text=n21c,fill=n21c!10},
n53cs/.style={draw=n53c,text=n53c,fill=n53c!10},
n30cs/.style={draw=n30c,text=n30c,fill=n30c!10},
n55cs/.style={draw=n55c,text=n55c,fill=n55c!10},
n40cs/.style={draw=n40c,text=n40c,fill=n40c!10},
n79cs/.style={draw=n79c,text=n79c,fill=n79c!10},
n64cs/.style={draw=n64c,text=n64c,fill=n64c!10},
n70cs/.style={draw=n70c,text=n70c,fill=n70c!10},
n13cs/.style={draw=n13c,text=n13c,fill=n13c!10},
scale=\tikzscale]
\node[nodified, n55cs] (n55)  at (0.710, -3.000){};
\node[nodified, n70cs] (n70)  at (0.740, -1.120){};
\node[nodified, n65cs] (n65)  at (3.410, -3.010){};
\node[nodified, n53cs] (n53)  at (1.060, -3.950){};
\node[nodified, n26cs] (n26)  at (0.910, -7.550){};
\node[node] (n22)  at (4.310, -7.560) {\ml{\textbf{End}}};
\node[node] (n17)  at (4.555, -1.210) {\ml{\textbf{Const}\\value = 1}};
\node[nodified, n64cs] (n64)  at (2.800, -2.010){};
\node[node, n15cs] (n15)  at (3.795, -4.350) {\ml{\textbf{Phi}}};
\node[nodified, n37cs] (n37)  at (0.370, -4.100){};
\node[node, n14cs] (n14)  at (1.545, -6.190) {\ml{\textbf{Block}}};
\node[nodified, n36cs] (n36)  at (2.080, -6.330){};
\node[nodified, n31cs] (n31)  at (4.540, -6.530){};
\node[nodified, n44cs] (n44)  at (2.260, -5.670){};
\node[nodified, n51cs] (n51)  at (2.330, -3.970){};
\node[node] (n19)  at (2.790, -7.050) {\ml{\textbf{Add}}};
\node[node, n4cs] (n4)  at (2.355, -4.570) {\ml{\textbf{Block}}};
\node[nodified, n23cs] (n23)  at (2.980, -7.540){};
\node[nodified, n62cs] (n62)  at (1.790, -2.000){};
\node[nodified, n28cs] (n28)  at (0.910, -6.540){};
\node[nodified, n30cs] (n30)  at (1.840, -7.030){};
\node[nodified, n67cs] (n67)  at (4.040, -2.660){};
\node[nodified, n46cs] (n46)  at (0.960, -5.670){};
\node[node, n2cs] (n2)  at (0.855, -1.760) {\ml{\textbf{Block}}};
\node[node] (n20)  at (1.060, -7.150) {\ml{\textbf{Return}}};
\node[nodified, n59cs] (n59)  at (0.980, -2.060){};
\node[node] (n10)  at (1.890, -3.400) {\ml{\textbf{LESS}}};
\node[nodified, n40cs] (n40)  at (4.400, -0.510){};
\node[nodified, n47cs] (n47)  at (2.210, -4.790){};
\node[node, n13cs] (n13)  at (2.355, -5.300) {\ml{\textbf{Jmp}}};
\node[node, n11cs] (n11)  at (1.805, -3.910) {\ml{\textbf{Cond}}};
\node[node, n12cs] (n12)  at (1.145, -5.330) {\ml{\textbf{Jmp}}};
\node[node] (n6)  at (2.855, -1.210) {\ml{\textbf{Const}\\value = 1}};
\node[nodified, n79cs] (n79)  at (1.070, -0.300){};
\node[node, n0cs] (n0)  at (2.090, -0.170) {\ml{\textbf{StartBlock}}};
\node[node] (n1)  at (0.525, -0.700) {\ml{\textbf{Start}}};
\node[node, n21cs] (n21)  at (2.005, -7.550) {\ml{\textbf{EndBlock}}};
\node[node] (n5)  at (1.795, -1.150) {\ml{\textbf{Const}\\value = 0}};
\node[nodified, n49cs] (n49)  at (1.010, -4.790){};
\node[nodified, n42cs] (n42)  at (3.050, -5.600){};
\node[nodified, n77cs] (n77)  at (1.760, -0.430){};
\node[nodified, n75cs] (n75)  at (2.580, -0.450){};
\node[nodified, n57cs] (n57)  at (2.410, -3.650){};
\node[nodified, n33cs] (n33)  at (3.340, -6.330){};
\node[node] (n9)  at (2.060, -2.650) {\ml{\textbf{Cmp}\\associative = false\\commutative = false}};
\node[node, n3cs] (n3)  at (1.115, -4.490) {\ml{\textbf{Block}}};
\path[edge, -] (n9)  -- node[lab]{$-$1} (n59) ;
\path[edge, -, n15cs] (n15)  -- node[lab]{$-$1} (n42) ;
\path[edge, n44cs] (n44)  -- (n13.south -| 2.260, -5.670);
\path[edge, n57cs] (n57)  -- (n9.south -| 2.410, -3.650);
\path[edge, n65cs] (n65)  --  (n5) ;
\path[edge, n64cs](n64.north -| 2.855, -1.210) --  (n6) ;
\path[edge, -] (n20)  -- (0.410, -6.620) -- (n37.south -| 0.410, -6.620);
\node[lab] at (0.424, -5.928){0};
\path[edge, -, n21cs](n21.west |- 0.910, -7.550) -- node[lab]{0} (n26) ;
\path[edge, n59cs] (n59)  -- (n2.south -| 0.980, -2.060);
;
\path[edge, -, n11cs] (n11)  -- node[lab]{$-$1} (n55) ;
\path[edge, n30cs](n30.east |- 2.790, -7.050) --  (n19) ;
\path[edge, -] (n19)  -- node[lab]{0} (n33) ;
\path[edge, -, n14cs] (n14)  -- node[lab]{1} (n46) ;
\path[edge, n47cs] (n47)  -- (n4.south -| 2.210, -4.790);
\path[edge, -](n22.west |- 2.980, -7.540) -- node[lab]{$-$1} (n23) ;
\path[edge, n53cs](n53.east |- 1.805, -3.910) --  (n11) ;
\path[edge, n62cs](n62.north -| 1.795, -1.150) --  (n5) ;
\path[edge, n75cs] (n75)  --  (n0) ;
\path[edge, n26cs] (n26)  -- (n20.south -| 0.910, -7.550);
\path[edge, n49cs] (n49)  -- (n3.south -| 1.010, -4.790);
\path[edge, -, n3cs](n3.north -| 1.060, -3.950) -- node[lab]{0} (n53) ;
\path[edge, -](n17.north -| 4.400, -0.510) -- node[lab]{$-$1} (n40) ;
\path[edge, -] (n19)  -- node[lab]{1} (n31) ;
\path[edge, -, n15cs] (n15)  -- node[lab]{1} (n65) ;
\path[edge, n46cs] (n46)  -- (n12.south -| 0.960, -5.670);
\path[edge, n67cs] (n67)  --  (n6) ;
\path[edge, n31cs](n31.north -| 4.555, -1.210) --  (n17) ;
\path[edge, -](n20.north -| 0.910, -6.540) -- node[lab]{$-$1} (n28) ;
\path[edge, n40cs] (n40)  --  (n0) ;
\path[edge, n28cs] (n28)  --  (n14) ;
\path[edge, n70cs] (n70)  -- (n1.south -| 0.740, -1.120);
\path[edge, n77cs] (n77)  -- (n0.south -| 1.760, -0.430);
\path[edge, -, n2cs](n2.north -| 0.740, -1.120) -- node[lab]{0} (n70) ;
\path[edge, -](n9.north -| 1.790, -2.000) -- node[lab]{0} (n62) ;
\path[edge, -] (n19)  -- node[lab]{$-$1} (n36) ;
\path[edge, n79cs] (n79)  -- (n0.west |- 1.070, -0.300);
\path[edge, n55cs] (n55)  -- (n2.south -| 0.710, -3.000);
\path[edge, -](n20.east |- 1.840, -7.030) -- node[lab]{1} (n30) ;
\path[edge, -, n12cs](n12.north -| 1.010, -4.790) -- node[lab]{$-$1} (n49) ;
\path[edge, n36cs] (n36)  -- (n14.east |- 2.080, -6.330);
\path[edge, n37cs] (n37)  -- (n1.south -| 0.370, -4.100);
\path[edge, -] (n9)  -- node[lab]{1} (n64) ;
\path[edge, -] (n1)  -- node[lab]{$-$1} (n79) ;
\path[edge, n23cs](n23.west |- 2.005, -7.550) --  (n21) ;
\path[edge, -, n4cs](n4.north -| 2.330, -3.970) -- node[lab]{0} (n51) ;
\path[edge, n33cs] (n33)  --  (n15) ;
\path[edge, -, n15cs] (n15)  -- node[lab]{0} (n67) ;
\path[edge, n51cs](n51.west |- 1.805, -3.910) --  (n11) ;
\path[edge, -](n5.north -| 1.760, -0.430) -- node[lab]{$-$1} (n77) ;
\path[edge, -, n11cs] (n11)  -- node[lab]{0} (n57) ;
\path[edge, -, n13cs](n13.north -| 2.210, -4.790) -- node[lab]{$-$1} (n47) ;
\path[edge, -, n14cs] (n14)  -- node[lab]{0} (n44) ;
\path[edge, -](n6.north -| 2.580, -0.450) -- node[lab]{$-$1} (n75) ;
\path[edge](n9.south -| 1.890, -3.400) -- node[lab]{relation} (n10) ;
\path[edge, n42cs] (n42)  --  (n14) ;
\userdefinedmacro
\end{tikzpicture}
\renewcommand{\userdefinedmacro}{\relax}

%% file: figs/consts.tikz
%
\begin{tikzpicture}[
scale=\tikzscale]
\node[node] (n6)  at (1.455, -1.155) {\ml{\textbf{Dataflow}\\position ?= $-$1}};
\node[node] (n0)  at (1.405, -0.255) {\ml{{\color{\red}\textbf{! StartBlock}}\\\textbf{Block}}};
\node[node] (n1)  at (1.405, -1.955) {\ml{\textbf{Const}}};
\path[edge](n6.north -| 1.405, -0.255) -- node[lab]{out} (n0) ;
\path[edge](n1.north -| 1.455, -1.155) -- node[lab]{in} (n6) ;
\userdefinedmacro
\end{tikzpicture}
\renewcommand{\userdefinedmacro}{\relax}

%% file: figs/add-fold.tikz
%
\begin{tikzpicture}[
scale=\tikzscale]
\node[node] (n6)  at (1.905, -2.680) {\ml{\textbf{\textit{Co1 : Const}}}};
\node[node] (n12)  at (1.220, -0.230) {\ml{\textbf{StartBlock}}};
\node[delnode] (n0)  at (1.170, -4.140) {\ml{\textbf{Add}}};
\node[delnode] (n1)  at (0.625, -3.370) {\ml{\textbf{Dataflow}\\position ?= 0}};
\node[node] (n5)  at (0.625, -2.690) {\ml{\textbf{\textit{Co0 : Const}}}};
\node[quantnode] (n16)  at (3.445, -3.305) {\ml{$\forall^{>0}$}};
\node[node] (n15)  at (2.905, -4.190) {\ml{\textbf{Dataflow}}};
\node[newnode] (n13)  at (1.300, -0.920) {\ml{\textbf{Dataflow}\\position := $-$1}};
\node[newnode] (n11)  at (2.165, -1.860) {\ml{\textbf{Const}\\value := Co0.value $+$ Co1.value}};
\node[delnode] (n2)  at (1.955, -3.390) {\ml{\textbf{Dataflow}\\position ?= 1}};
\path[newedge](n11.north -| 1.300, -0.920) -- node[newlab]{in} (n13) ;
\path[deledge](n1.north -| 0.625, -2.690) -- node[dellab]{out} (n5) ;
\path[deledge](n2.north -| 1.905, -2.680) -- node[dellab]{out} (n6) ;
\path[deledge](n15.west |- 1.170, -4.140) -- node[dellab]{out} (n0) ;
\path[deledge] (n0)  -- node[dellab]{in} (n2) ;
\path[newedge] (n15)  -- node[newlab]{out}(n11.south -| 2.905, -4.190);
\path[newedge](n13.north -| 1.220, -0.230) -- node[newlab]{out} (n12) ;
\path[quantedge] (n15)  -- node[lab]{at} (n16) ;
\path[deledge] (n0)  -- node[dellab]{in} (n1) ;
\userdefinedmacro
\end{tikzpicture}
\renewcommand{\userdefinedmacro}{\relax}

%% file: main.bbl
\begin{thebibliography}{1}
\providecommand{\bibitemdeclare}[2]{}
\providecommand{\urlprefix}{Available at }
\providecommand{\url}[1]{\texttt{#1}}
\providecommand{\href}[2]{\texttt{#2}}
\providecommand{\urlalt}[2]{\href{#1}{#2}}
\providecommand{\doi}[1]{doi:\urlalt{http://dx.doi.org/#1}{#1}}
\providecommand{\bibinfo}[2]{#2}

\bibitemdeclare{inproceedings}{compileroptimizationcase}
\bibitem{compileroptimizationcase}
\bibinfo{author}{Sebastian Buchwald} \& \bibinfo{author}{Edgar Jakumeit}
  (\bibinfo{year}{2011}): \emph{\bibinfo{title}{Compiler Optimization: A Case
  for the Transformation Tool Contest}}.
\newblock In \bibinfo{editor}{Pieter {Van Gorp}}, \bibinfo{editor}{Steffen
  Mazanek} \& \bibinfo{editor}{Louis Rose}, editors: {\sl
  \bibinfo{booktitle}{{TTC} 2011: Fifth Transformation Tool Contest, Z\"urich,
  Switzerland, June 29-30 2011}}, \bibinfo{publisher}{{EPTCS}}.

\bibitemdeclare{article}{sttt}
\bibitem{sttt}
\bibinfo{author}{A.~Ghamarian}, \bibinfo{author}{M.~de~Mol},
  \bibinfo{author}{A.~Rensink}, \bibinfo{author}{E.~Zambon} \&
  \bibinfo{author}{M.~Zimakova} (\bibinfo{year}{2011}):
  \emph{\bibinfo{title}{Modelling and analysis using \GROOVE}}.
\newblock {\sl \bibinfo{journal}{International Journal on Software Tools for
  Technology Transfer (STTT).}} \doi{10.1007/s10009-011-0186-x}.

\bibitemdeclare{misc}{share}
\bibitem{share}
\emph{\bibinfo{title}{{SHARE} demo related to the paper: Solving the TTC 2011
  Compiler Optimization Case with {GROOVE}}}.
\newblock
  \bibinfo{note}{\url{http://is.ieis.tue.nl/staff/pvgorp/share/?page=Configure%
NewSession&vdi=Ubuntu-11_TTC11_groove-cop.vdi}}.

\end{thebibliography}
